\documentclass[letterpaper,12pt,double space]{revtex4}
\usepackage{graphicx}
\usepackage{dcolumn}
\usepackage{bm}

\begin{document}

\title{Influence of unsymmetrical periodicity on extraordinary transmission through periodic arrays of subwavelength holes}
\author{Xi-Feng Ren,Guo-Ping Guo\footnote[2]{gpguo@ustc.edu.cn}, Yun-Feng Huang,Zhi-Wei Wang,Guang-Can Guo}
\address{Key Laboratory of Quantum Information, University of Science and Technology of China, Hefei
230026, People's Republic of China}

\begin{abstract}
Quadrate hole array is explored to study the influence of
unsymmetrical periodicity on extraordinary optical transmission
through periodic arrays of subwavelength holes. It is found that the
transmission efficiency of light and the ratio between transmission
efficiencies of horizontal and vertical polarized light can be
continuously tuned by rotating the quadrate hole array. We can
calculate out the transmission spectra (including the heights and
locations of peaks) for any rotation angle $\theta$ with a simple
theoretical model.

PACS number(s): 78.66.Bz,73.20.MF, 71.36.+c
\end{abstract}

\maketitle

\newpage

The extraordinary optical transmission (EOT) through periodic arrays
of subwavelength holes has attracted much attention since it was
first reported in 1998\cite{1}. Generally, it is believed that metal
surface plays a crucial role and the phenomenon is mediated by
surface plasmons and there is a process of transforming photon to
surface plasmon and back to photon\cite{4,crucial,ebbesen5}. The
polarization of the incident light determines the mode of excited
surface plasmon which is also related to the periodic structure.
This phenomenon can be used in various applications, for example,
sensors, opto electronic device, etc\cite{williams, brolo, nahata,
luo, shinada}. Polarization properties of nanohole arrays have been
studied in many works\cite{Elli04,Gordon04,Altew05}. Recently,
orbital angular momentum of photons were explored to investigate the
spatial mode properties of surface plasmon assisted transmission
\cite{ren061,ren062}. The shape distortion of single wavepacket
during photon to plasmon and back to photon process was also
discussed\cite{ren063}. In 2002, E. Altewischer \textit{et al.}
\cite{alt} also showed that quantum entanglement of photon pairs can
be preserved when they respectively travel through a hole array.
Therefore, the macroscopic surface plasmon polarizations, a
collective excitation wave involving typically $10^{10}$ free
electrons propagating at the surface of conducting matter, have a
true quantum nature. However, the increasing use of EOT requires
further understanding of the phenomenon.

For the manipulation of light at a subwavelength scale with periodic
arrays of holes, two ingredients exist: shape and
periodicity\cite{4,crucial,ebbesen5,klein,elliott,Ruan}. In this
work, we used a quadrate hole array to investigate the influence of
unsymmetrical periodicity on EOT. It was found that the optical
transmission spectra were changed when we rotated the hole array.
The transmission spectra strongly depends on the rotation angle, in
other words, the angle between polarization of incident light and
axis of hole array. We also gave a simple model to explain the
results. Using this feature, we can continuously tune the
transmission efficiency of light with certain polarization by
rotating the quadrate hole array. The ratio between transmission
efficiencies of horizontal polarized and vertical polarized light
can also be modified.

Fig. 1 shows sketch maps of our hole arrays. (a) is a square hole
array. It is produced as follows: after subsequently evaporating a
$3$-$nm$ titanium bonding layer and a $135$-$nm$ gold layer onto a
$0.5$-$mm$-thick silica glass substrate, a Electron Beam Lithography
System (EBL) is used to produce cylindrical holes ($200nm$ diameter)
arranged as a square lattice ($600nm$ period). The area of the hole
array is $300\mu m\times 300\mu m$. (b) is a quadrate hole array. It
is produced by a Focused Ion Beam Etching system (FIB). Periods in
two axes are $600nm$ and $540nm$ respectively. The area of the hole
array is $42\mu m\times 44\mu m$.

Transmission spectra of the hole arrays were recorded by a Silicon
avalanche photodiode (APD) single photon counter couple with a
spectrograph through a fiber. White light from a stabilized
tungsten-halogen source passed though a single mode fiber and a
polarizer (only vertical polarized light can pass), then illuminated
the sample. The hole arrays were set between two lenses of 35
\emph{mm} focal length, so that the light was normally incident on
the hole array with a cross sectional diameter about $20\mu m$ and
covered hundreds of holes. The light exiting from the hole array was
launched into the spectrograph.

In our experiment, the hole arrays were rotated anti-clockwise in
the plane perpendicular to the illuminating light, as shown in Fig.1
(c). First, we used the square sample and recorded the transmission
spectra with rotation angle $\theta=0\textordmasculine ,
30\textordmasculine , 60\textordmasculine, 90\textordmasculine $
respectively. The results were shown in Fig. 2. We can see that
there is no obvious change among the spectra of four cases.

Then we used the quadrate hole array and recorded the transmission
spectra with rotation angle $\theta=0\textordmasculine ,
15\textordmasculine , 30\textordmasculine , 45\textordmasculine ,
60\textordmasculine, 75\textordmasculine ,90\textordmasculine $
respectively. The results were shown in Fig. 3. Much different from
the case of square sample, the spectra had a large change in both
the heights and the locations of peaks. For
$\theta=0\textordmasculine $, the wavelengths at two peaks were
about $555nm$ and $700nm$ with transmission efficiencies $1.21\%$
and $6.30\%$. While for $\theta=90\textordmasculine $, the
wavelengths at two peaks were about 570 nm and 680nm with
transmission efficiencies $1.53\%$ and $3.99\%$. For other rotation
angles, the heights and the locations were located between the upper
two cases. The transmission spectra were more close to the case of
$\theta=90\textordmasculine $ with increasing of $\theta$.

To explain this phenomenon, we gave a simple model. Since the
surface plasmons were excited in the directions of long (L) and
short (S) axes of quadrate hole array, we can suspect that this two
directions were eigenmode-directions for our sample. The
polarization of illuminating light was projected into the two
eigenmode-directions to excite surface plasmons. After that, the two
kinds of surface plasmons transmitted the holes and irritated light
with transmission efficiencies $T_{L}$ and $T_{S}$ respectively. So
for light whose polarization had an angle $\theta$ with the $L$
axis, the transmission efficiency $T_{\theta}$ will be

\begin{equation}
T_{\theta }=T_{L}\cos^2 (\theta)+T_{S}\sin^2 (\theta).
\end{equation}
This equation was appropriate for any wavelength. So if we know the
transmission spectra for enginmode-directions (here L and S), we can
calculate out the transmission spectra (including the heights and
locations of peaks) for any $\theta$. Fig. 4 gave the comparison
between the calculations (lines) and experimental results (dots) for
$\theta=15\textordmasculine , 30\textordmasculine ,
45\textordmasculine , 60\textordmasculine, 75\textordmasculine $. It
can be seen that the experimental data fit the lines well, which
sustains our model. The surface plasmon was highly polarization
dependent, so the polarization of transmitted light was composed of
two parts: directions along the L and S axes of quadrate hole array.
For rotation angle $\theta$, polarization $V$ was changed to
$(\sqrt{T_{L}}\cos^2(\theta)+\sqrt{T_{S}}\sin^2
(\theta))|V\rangle+(\sqrt{T_{S}}-\sqrt{T_{L}})\sin(\theta)\cos(\theta)|H\rangle$
(not normalized). Since the square hole array had the relation:
$T_{L}=T_{S}$ for any wavelength, transmission efficiencies will
always be $T_{\theta }=T_{L}=T_{S}$ and the transmission spectra
will not be influenced, as shown in Fig. 2. Unlike the square hole
array, not only the transmission efficiencies of light, but also the
ratio of transmission efficiencies between vertical and horizontal
polarized light were changed when we rotated the quadrate hole
array. For rotation angle $\theta$, the ratio was

\begin{equation}
R_{\theta }=T_{V_{\theta }}/T_{H_{\theta}}
=(T_{L}\cos^2(\theta)+T_{S}\sin^2 (\theta))/(T_{L}\sin^2
(\theta)+T_{S}\cos^2 (\theta)).
\end{equation}
As an example, the case for $700nm$ wavelength light was shown in
Fig. 5. Obviously, the ratio could be varied in a range by changing
the rotation angle $\theta$. Of course, the polarization of
transmitted light will be changed due to this unequal transmission
efficiencies. Our model may also be extended to other structures,
such as metal plate with subwavelength slits. While for different
structures and materials, the eigenmode-directions might be
different.

In conclusion, quadrate hole array was explored to study the
influence of unsymmetrical periodicity on EOT. Because of the
reduced symmetry of structure, the transmission spectra were changed
when we rotated the hole array. A simple model was given to explain
this phenomenon. Using this protocol, we could continuously tune the
transmission efficiency of light with certain polarization by
rotating the quadrate hole array. The ratio of transmission
efficiencies between horizontal and vertical polarized light could
also be modified. Our results may be useful in the further
application of plasmonics.

This work was funded by the National Fundamental Research Program,
National Nature Science Foundation of China (10604052), the
Innovation Funds from Chinese Academy of Sciences, the Program of
the Education Department of Anhui Province (Grant No.2006kj074A)and
China Postdoctoral Science Foundation(20060400205).

\newpage

\section*{List of Figure Captions}
Fig. 1. Sketch picture of our hole arrays: (a) Square hole array.
Period is 600nm. (b) Quadrate hole array. Periods are 600nm and
540nm in two directions respectively. (c) Rotation of our hole
arrays. S (L) is the axis of short (long) period of quadrate hole
array; H(V) is horizontal (vertical) axis.

Fig. 2. (Color online)Transmittance as a function of wavelength when
we rotated the square hole array. The rotation angle are
0\textordmasculine (black square dots), 30\textordmasculine (red
round dots), 60\textordmasculine (green triangle dots),
90\textordmasculine (blue inverse triangle dots) respectively.

Fig. 3. (Color online)Transmittance as a function of wavelength when
we rotated the quadrate hole array. The rotation angle are
0\textordmasculine (black square dots), 15\textordmasculine (red
round dots), 30\textordmasculine (green triangle dots),
45\textordmasculine (blue inverse triangle dots),
60\textordmasculine (cyan diamond dots), 75\textordmasculine
(magenta pentagon dots), 90\textordmasculine (yellow hexagon dots)
respectively.

Fig. 4. (Color online)Comparison between theoretical calculations
(lines) and experimental results (dots). Rotation angle: (a)
15\textordmasculine , (b) 30\textordmasculine , (c)
45\textordmasculine , (d) 60\textordmasculine , (e)
75\textordmasculine . The individual spectra are offset vertically
by 1.5\% for clarity. In all the cases, experimental data fit
theoretical calculations well.

Fig. 5. (Color online)Ratio of transmission efficiencies between
vertical and horizontal polarized light (wavelength $700nm$). Black
dots are experimental results.

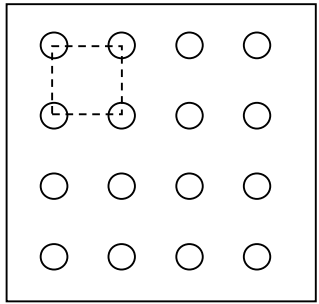
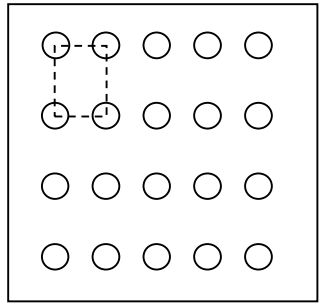
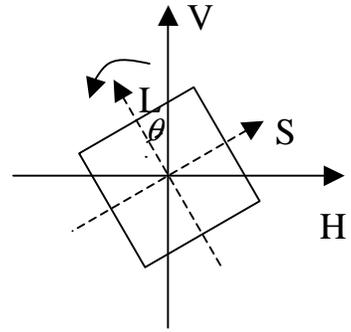
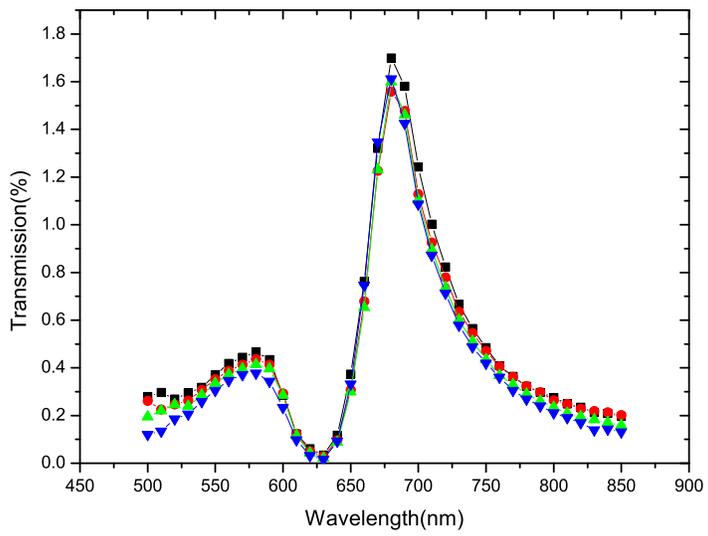
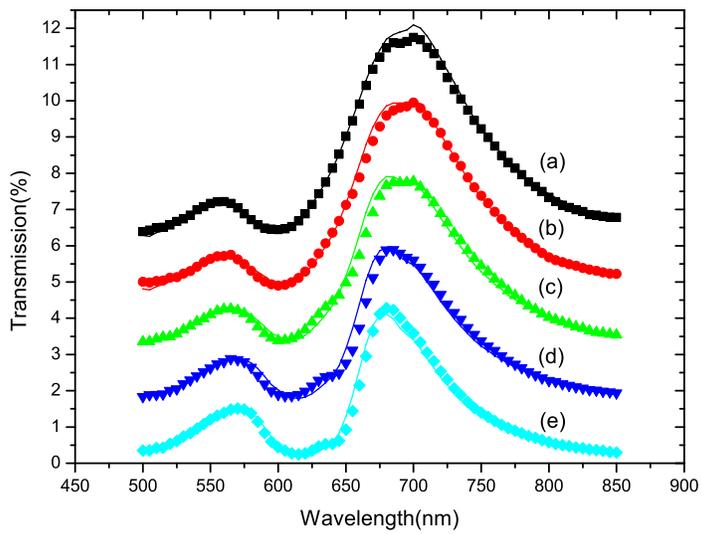
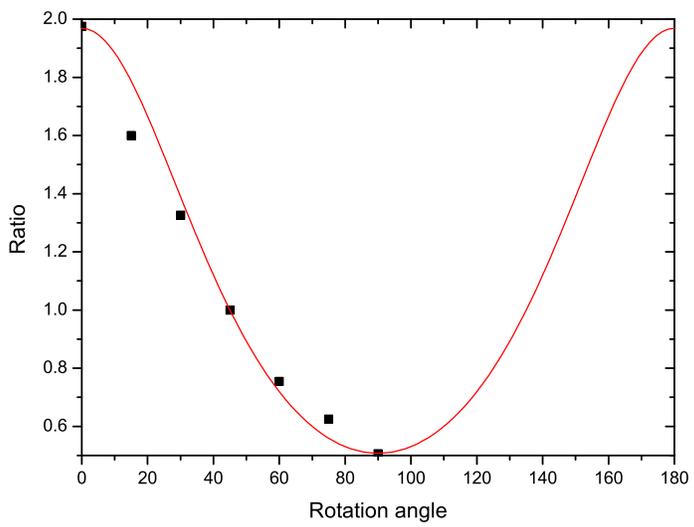
\end{document}